\documentclass{ws-mpla}

\begin{document}

\markboth{Gabriele Honecker}
{Chiral N=1 4d Orientifolds with D-branes at Angles}

%
\catchline{}{}{}{}{}
%

\title{CHIRAL N=1 4D ORIENTIFOLDS WITH D-BRANES AT ANGLES}

\author{\footnotesize GABRIELE HONECKER}

\address{Departamento de F\'{\i}sica Te\'orica C-XI and Instituto de F\'{\i}sica Te\'orica C-XVI,\\
Universidad Aut\'onoma de Madrid, Cantoblanco, 28049 Madrid,
Spain\\
gabriele@th.physik.uni-bonn.de}

\maketitle

\pub{Received 15 June 2004}{}

\begin{abstract}
D6-branes intersecting at angles allow for phenomenologically 
appealing constructions of four dimensional string theory vacua.
While it is straightforward to obtain non-supersymmetric realizations
of the standard model, supersymmetric and stable models with
three generations and no exotic chiral matter require more involved orbifold constructions.
The $T^6/({\mathbb Z}_4 \times {\mathbb Z}_2 \times \Omega {\cal R})$ case is 
discussed in detail. Other orbifolds including fractional D6-branes are treated briefly.

\end{abstract}

\ccode{PACS Nos.: 11.25.Mj, 12.60.Jv, 11.25.Uv.}

\section{Introduction}  

Intersecting D6-branes at angles\cite{Berkooz:1996km,Blumenhagen:2000wh,Aldazabal:2000dg,Aldazabal:2000cn}
(for a more complete list of references see\cite{Uranga:2003pz,Lust:2004ks,Kiritsis:2003mc,Angelantonj:2002ct})\footnote{For works in the T-dual picture with D9-branes and magnetic backgrounds see e.g.~\cite{Bachas:1995ik,Angelantonj:2000hi,Angelantonj:2000rw,Larosa:2003mz}.}
provide a powerful framework for model building in type II superstring 
backgrounds. While compactifications on generic Calabi-Yau manifolds only allow to 
compute the RR tadpoles and the chiral spectrum, in toroidal and orbifold backgrounds
conformal field theory methods facilitate the computation of the complete spectrum, and
interactions have a geometric interpretation
along the compact dimensions\cite{Aldazabal:2000cn}.
The Yukawa\cite{Cvetic:2002wh,Cremades:2003qj,Cvetic:2003ch,Abel:2003vv,Cremades:2004wa,Lust:2004cx} 
and n-point couplings\cite{Abel:2003yx}, proton decay\cite{Klebanov:2003my},
flavor changing neutral
currents\cite{Abel:2003fk,Abel:2003yh}, threshold corrections\cite{Lust:2003ky} as well as the 
tree-level K\"ahler- and
superpotential\cite{Lust:2004cx} can be explicitly computed in this framework.

Supersymmetric D6-branes at angles  arise naturally in four dimensional orbifold/orientifold compactifications of 
type IIA string theory\cite{Blumenhagen:1999ev,Forste:2000hx,Blumenhagen:2002wn,Blumenhagen:2004di},
\footnote{For the first chiral supersymmetric spectra in this framework see also~\cite{Cvetic:2001tj,Cvetic:2001nr}.} 
if the worldsheet parity $\Omega$ is accompanied by an involution
${\cal R}$ on the six compact dimensions, i.e. ${\cal R}^2=1$, which leaves half of the internal directions  
invariant. The fix loci of ${\cal R}\theta^k$, where $\theta$ generates the orbifold background, 
support O6-planes extended along
${\mathbb R}^{1,3}$ plus three compact dimensions,
whose charge under the RR 7-form is canceled by the addition of D6-branes also wrapping ${\mathbb R}^{1,3}$ 
and three compact dimensions. In general, the D6-branes do not wrap the ${\cal R}$ invariant cycles.
For ${\cal R}$ to be a symmetry, therefore, brane images have to be included.
If the O6-planes wrap ${\mathbb R}^{1,3} \times \Pi_{O6}$, the D$6_a$-branes extend along  ${\mathbb R}^{1,3} \times \Pi_{a}$     
and their ${\cal R}$ images along ${\mathbb R}^{1,3} \times \Pi_{a'}$ with $\Pi_{a'} \equiv {\cal R} (\Pi_{a})$ and 
$\Pi_{O6}, \Pi_{a}, \Pi_{a'}$ denoting compact 3-cycles. The condition for the cancellation of the RR 7-form charge reads
\begin{equation} 
\sum_a \left( \int_{{\mathbb R}^{1,3} \times \Pi_{a}}C_7 + \int_{{\mathbb R}^{1,3} \times \Pi_{a'}}C_7 \right)
+ Q_{O6}\int_{{\mathbb R}^{1,3} \times \Pi_{O6}} C_7 =0. \nonumber
\end{equation}
Including the possibility of $N_a$ identical D$6_a$-branes and inserting the value of the charge of the O6-plane $Q_{O6}=-4$ leads then to the 
following condition on the compact 3-cycles
\begin{equation} \label{Eq:Tadp_general}
\sum_a N_a \left( \Pi_a + \Pi_{a'} \right) - 4 \Pi_{O6}=0,
\end{equation}
i.e. the cancellation of RR tadpoles is the Poincar\'{e} dual of  
 the requirement of an overall vanishing class of homology 3-cycles.

The chiral spectrum as well as the RR tadpole cancellation can be expressed
 purely in terms of the 
topological setting along the compact dimensions.
The intersection number $\Pi_a \circ \Pi_b$ between two  kinds of D$6_a$- and D$6_b$-branes counts the net
 number of chiral bifundamental representations 
of the gauge groups $U(N_a) \times U(N_b)$ involved. These bifundamental states are located at the 
pointlike singularity arising at the intersection of the 3-cycles 
along the compact dimensions. A negative intersection number signals the existence of the complex conjugate representation.\cite{Blumenhagen:2000wh}
The generic chiral spectrum is listed in table~1.
\begin{table}[h]
\tbl{Generic chiral spectrum} 
{\begin{tabular}{@{}cc@{}} \toprule
multiplicity & rep.\\\colrule
$\Pi_a \circ \Pi_b$ &  $({\bf N}_a,\overline{\bf N}_b)$ \\
$\Pi_a \circ \Pi_{b'}$ &   $({\bf N}_a,{\bf N}_b)$\\
$\frac{1}{2}\left(\Pi_a \circ \Pi_{a'}-\Pi_a \circ \Pi_{O6} \right)$ &  ${\bf Sym}_a$ \\
$\frac{1}{2}\left(\Pi_a \circ \Pi_{a'}+\Pi_a \circ \Pi_{O6} \right)$
&  ${\bf Anti}_a$\\\botrule
\end{tabular}}
\end{table}

In generic compactifications with $N_a$  D$6_a$-branes wrapping cycles $\Pi_a \neq \Pi_{a'}$,  
unitary gauge groups $U(N_a)$ arise which decompose then as $SU(N_a) \times U(1)_a$.
For factorizable cycles with 
$\Pi_a = \Pi_{a'}$, additional open string states become massless, and $N_a$ branes plus their ${\cal R}$ images support 
the gauge group $Sp(2N_a)$ or $SO(2N_a)$.
The cubic non-Abelian $SU(N_a)^3$
gauge anomalies vanish automatically upon RR tadpole cancellation~(\ref{Eq:Tadp_general}),
whereas mixed anomalies $SU(N_a)^2 - U(1)_b$ and $U(1)_a^2 - U(1)_b$ are in general present.
The anomalous $U(1)$ factors acquire a mass through the generalized Green Schwarz mechanism.
The relevant couplings to four dimensional RR scalars and their Hodge dual 2-forms
are computed from the geometry of the compact space as follows.\cite{Cremades:2002cs,MarchesanoBuznego:2003hp,Ott:2003yv}
Instead of expressing the 3-cycles in terms of an arbitrary basis, two sets of linearly independent cycles $\Pi_i^{\pm}$
can be chosen such that ${\cal R} \left(\Pi_i^{\pm}\right) = \pm \Pi_i^{\pm}$ with 
$\Pi_i^{+} \circ \Pi_j^{-} = c \delta_{ij}$ and $c$ a model dependent constant.
The four dimensional RR scalars and 2-forms then arise as the pullbacks of the $\Omega$ even 
RR 3-form over ${\cal R}$ even 3-cycles and the $\Omega$ odd 5-form over ${\cal R}$ odd 3-cycles,
\begin{equation}\label{Eq:4dRRfields}
\phi_i = (4\pi^2\alpha')^{-3/2} \int_{\Pi_i^{+}} {}^{(10)}C_3, \qquad
B^i_2 = (4\pi^2\alpha')^{-3/2}  \int_{\Pi_j^{-}} {}^{(10)}C_5.
\end{equation}
A general 3-cycle and its ${\cal R}$ image can be expanded in terms of its ${\cal R}$ even and odd components,
\begin{equation}
\Pi_a = \sum_i \left(r_a^i \Pi_i^{+} + s_a^i \Pi_i^{-}\right), \qquad 
\Pi_{a'} = \sum_i \left(r_a^i \Pi_i^{+} - s_a^i \Pi_i^{-}\right).  \nonumber
\end{equation}
The expansion coefficients $r^i_a$, $s^i_a$ of the 3-cycles reduce in the effective four dimensional theory  
to the coefficients of the Green Schwarz couplings,
\begin{equation}\label{Eq:GScouplings}
\sum_{i} 2r^i_b  \int_{{\mathbb R}^{1,3}}  \phi_i \;\mbox{tr} F_b \wedge F_b, \qquad
 N_a \sum_{i}2 s^i_a  \int_{{\mathbb R}^{1,3}} B^i_2 \wedge \mbox{tr} F_a.
\end{equation}
The latter coupling induces a mass of the Abelian gauge field involved.
Therefore, the massless $U(1)$ factors $Q=\sum_a x_a Q_a$ are those 
with a vanishing coupling to the 2-forms,
\begin{equation}\label{Eq:masslessAbelian}
\sum_a x_a N_a \vec{s}_a=0.
\end{equation}

\section{The $T^6/({\mathbb Z}_4 \times {\mathbb Z}_2 \times \Omega {\mathcal R})$ Orientifold}

\subsection{Geometry, supersymmetry and RR tadpole cancellation}\label{Subsec:Geometry}

The  orbifold  $T^6/({\mathbb Z}_4 \times {\mathbb Z}_2)$ has two generators,
\begin{eqnarray}
\Theta: \qquad &(z^1,z^2,z^3) \rightarrow (iz^1,-iz^2,z^3),\nonumber\\
\omega: \qquad &(z^1,z^2,z^3) \rightarrow (z^1,-z^2,-z^3),\nonumber
\end{eqnarray}
where $z^k=x^{2+2k}+ix^{3+2k}$ label the internal complex coordinates and $x^0, \ldots, x^3$ are 
the non-compact ones.
The involution which accompanies the worldsheet parity is chosen to be
\begin{equation}
{\cal R}: z^k \longrightarrow \overline{z}^{\overline{k}}, \qquad k=1,2,3.     \nonumber
\end{equation}
For consistency of the compactification, the orbifold lattice has to be invariant under
this involution. The square tori $T^2_1$, $T^2_2$ required by the 
${\mathbb Z}_4$ symmetry  have two possible orientations {\bf A} and {\bf B} 
w.r.t. the ${\cal R}$ invariant axes $x^{2k+2}$. The geometry is depicted in 
figure~1. The complex structure on $T^2_3$ is not fixed by the ${\mathbb Z}_2$ 
symmetry, and the torus can be either rectangular with the axes along $x^8$
and $x^9$ or tilted as depicted.\cite{Blumenhagen:2000ea}
\begin{figure}[th]
\centerline{\psfig{file=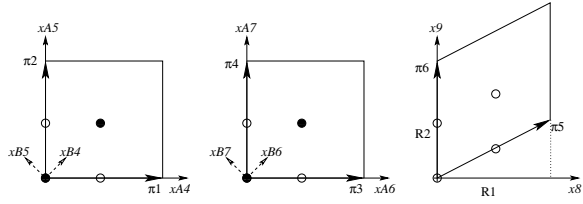,width=3.0in}}
\vspace*{8pt}
\caption{Geometric set-up of the $T^6/({\mathbb Z}_4 \times {\mathbb Z}_2 \times \Omega {\mathcal R})$ orientifold. 
On $T^2_1 \times T^2_2$, the 
square tori can have either orientation {\bf A} or {\bf B} w.r.t. the ${\cal R}$ invariant axis $x^{2+2k}$ ($k=1,2$). 
The torus $T^2_3$ can be either
untilted or tilted as depicted. Empty circles denote ${\mathbb Z}_2$ fixed
points while filled circles correspond to points
fixed under ${\mathbb Z}_4$. For further details see the paragraph below eq.~(\ref{Eq:Def_YZ}).}
\end{figure}

The Hodge numbers of this orbifold are $h_{1,1}=61, h_{2,1}=1$.\cite{Klein:2000qw}
This means, that $b_3=2+2h_{2,1}=4$ linearly independent 3-cycles exist.
As in the $T^6/({\mathbb Z}_2 \times {\mathbb Z}_2)$ case\cite{}, all 3-cycles are inherited from the underlying
torus and are given by orbifold invariant quantities, e.g.
 $\rho^{\prime}_1 =( \sum_{k=0}^3\Theta^k ) (\sum_{l=0}^1\omega^l )\pi_{135}$
 with $\pi_{135} \equiv \pi_1 \otimes \pi_3 \otimes \pi_5$.
 A convenient choice of linearly independent 3-cycles is
\begin{eqnarray}
\rho^{\prime}_1 &=4\left(\pi_{135}-\pi_{245}\right),
\qquad \rho^{\prime}_3 =4\left(\pi_{136}-\pi_{246}\right),\nonumber \\
\rho^{\prime}_2 &=4\left(\pi_{235}+\pi_{145}\right),
\qquad  \rho^{\prime}_4 =4\left(\pi_{236}+\pi_{146}\right),\nonumber
\end{eqnarray}
where the factor 4 arises from the invariance of any cycle under $\Theta^2$ and $\omega$, and e.g. $\Theta(\pi_{135})= -\pi_{245}$.
As in the $T^6/({\mathbb Z}_2 \times {\mathbb Z}_2)$ case (which has eight 
independent 3-cycles), the unimodular basis is formed by 
the cycles $\rho_i \equiv \frac{1}{2} \rho^{\prime}_i$ with intersection matrix
\begin{equation}\label{IntersectionMatrix}
I^{{\mathbb Z}_4 \times {\mathbb Z}_2} = \left(\begin{array}{cccc}
0 & 0 & -1 & 0\\
0 & 0 & 0 & -1\\
1 & 0 & 0 & 0 \\
0 & 1 & 0 & 0
\end{array}\right).
\end{equation}
Any  3-cycle can be expressed as a linear combination of the $\rho_i$. The coefficients of the factorizable 
3-cycles arise as follows. A factorizable cycle $a$ and its ${\mathbb Z}_4$ image $(\Theta a)$ are specified by the 
wrapping numbers $(n^a_k,m^a_k)$ along the 1-cycles $(\pi_{2k-1},\pi_{2k})$, 
\begin{eqnarray}
a: &\qquad& \left(n^a_1 \pi_1 + m^a_1 \pi_2\right) \otimes \left(n^a_2 \pi_3 + m^a_2 \pi_4\right) \otimes 
\left(n^a_3 \pi_5 + m^a_3 \pi_6\right), \nonumber \\
(\Theta a): &\qquad& \left(-m^a_1 \pi_1 + n^a_1 \pi_2\right) \otimes \left(m^a_2 \pi_3 - n^a_2 \pi_4\right) \otimes 
\left(n^a_3 \pi_5 + m^a_3 \pi_6\right).
\end{eqnarray}
The homology 3-cycle wrapped by a factorizable D$6_a$-brane therefore is
\begin{equation}\label{cycle_YZ}
\Pi_a = Y_a n^a_3  \rho_1 +  Z_a n^a_3 \rho_2 + Y_a m^a_3 \rho_3 + Z_a m^a_3 \rho_4,
\end{equation}
where the coefficients are ${\mathbb Z}_4$ invariant combinations of the wrapping numbers, 
\begin{equation}\label{Eq:Def_YZ}
Y_a = \left(n^a_1 n^a_2 - m^a_1 m^a_2 \right), \qquad
Z_a = \left(n^a_1 m^a_2 + m^a_1 n^a_2 \right).
\end{equation}

The ${\cal R}$ image cycle $\Pi_{a'}$ differs for the six inequivalent choices of lattice orientations:
on $T^2_1 \times T^2_2$, these are {\bf AA}, {\bf AB} and {\bf BB};
on $T^2_3$, a rectangular torus {\bf a} with arbitrary ratio of radii $r \equiv R_2/R_1$ has the ${\cal R}$ 
invariant axis $\pi_5$, and on a tilted torus {\bf b} with $r$ arbitrary 
$(2\pi_5 - \pi_6)$ is ${\cal R}$  invariant.
The latter two choices can be compactly treated by defining $b=0,\frac{1}{2}$ for the rectangular and tilted torus,
respectively. The specific choices $r=\frac{1}{1-b}$ correspond to square tori with orientations {\bf A} and {\bf B} 
for $b=0,\frac{1}{2}$, respectively. Observe that the choice of basis of the
{\bf B} orientation on $T^2_3$ differs from those on $T^2_1 \times T^2_2$.

The ${\cal R}$ images of the basis follow from tensoring the results of the
1-cycles, in detail
$\pi_{2k-1}\stackrel{\cal R}{\rightarrow}\pi_{2k-1}$, $\pi_{2k}\stackrel{\cal R}{\rightarrow}-\pi_{2k}$ per {\bf A} and
$\pi_{2k-1} \stackrel{\cal R}{\leftrightarrow} \pi_{2k}$ per {\bf B} torus ($k=1,2$) and 
$\pi_5 \stackrel{\cal R}{\rightarrow}\pi_5 -(2b) \pi_6$, 
$\pi_6\stackrel{\cal R}{\rightarrow} -\pi_6$ on $T^2_3$. The result is listed in table~2. 
\begin{table}[h]
\tbl{${\cal R}$ images of cycles for $T^6/({\mathbb Z}_4 \times {\mathbb Z}_2 \times \Omega{\cal R})$}
{\begin{tabular}{@{}ccccc@{}} \toprule
lattice  &  ${\cal R}:\rho_1$ &  ${\cal R}:\rho_2$  &  ${\cal R}:\rho_3$ &  ${\cal R}:\rho_4$ \\ \colrule
{\bf AAa/b} & $\rho_1-(2b) \rho_3$ & $-\rho_2+(2b) \rho_4$ & $-\rho_3$ & $\rho_4$  \\
{\bf ABa/b} & $\rho_2-(2b) \rho_4$ & $\rho_1-(2b) \rho_3$ & $-\rho_4$ & $-\rho_3$  \\
{\bf BBa/b} & $-\rho_1+(2b) \rho_3$ & $\rho_2-(2b) \rho_4$  & $\rho_3$ & $-\rho_4$ 
 \\\botrule
\end{tabular}}
\end{table}

The ${\cal R}$ image of the cycle $\Pi_a$ for the different lattice choices is given by 
\begin{equation}\label{Eq:Rimagecycles}
\Pi_{a'}= \left\{ \begin{array}{ll}
Y_a n^a_3 \rho_1-Z_a n^a_3 \rho_2-Y_a \left[m^a_3+(2b)n^a_3\right]\rho_3+ Z_a\left[m^a_3+(2b)n^a_3\right]\rho_4 
& {\bf AAa/b},\\
Z_a n^a_3 \rho_1+Y_a n^a_3 \rho_2-Z_a \left[m^a_3+(2b)n^a_3\right]\rho_3-Y_a\left[m^a_3+(2b)n^a_3\right]\rho_4 
 & {\bf ABa/b},\\
-Y_a n^a_3 \rho_1+Z_a n^a_3 \rho_2+Y_a \left[m^a_3+(2b)n^a_3\right]\rho_3- Z_a\left[m^a_3+(2b)n^a_3\right]\rho_4 
 & {\bf BBa/b}.
\end{array}\right.
\end{equation}

An arbitrary cycle $\Pi_a$ at non-trivial angles w.r.t. the ${\cal R}$ invariant axis on a generic $T^2_3$ does not 
preserve supersymmetry. The supersymmetry condition $\sum_{k=1}^3 \pi\tilde{\varphi}^a_k=0$,\cite{Berkooz:1996km} --- where 
$\pi\tilde{\varphi}^a_k= \pi(\varphi^a_k-\frac{c}{4})$ ($c=0,1$ for {\bf A}, {\bf B},
respectively, and $k=1,2$) is the angle
w.r.t. the ${\cal R}$ invariant plane on $T^2_k$ and $\pi\varphi^a_k$ is the angle w.r.t. $\pi_{2k-1}$ ---  is
cumbersome for computational purposes and more conveniently replaced by two
conditions on the wrapping numbers $n^a_3, m^a_3$
and coefficients $Y_a, Z_a$ by using the fact that a vanishing sum over angles implies 
\begin{equation}
\sum_{i=1}^{3}  \tan \pi \tilde{\varphi}^a_i = \prod_{i=1}^{3}  \tan \pi \tilde{\varphi}^a_i. 
\end{equation}
Due to 
\begin{equation}
T^2_i: i=1,2  \qquad   \tan \pi \varphi^a_i =\frac{m^a_i}{n^a_i}, \qquad \qquad 
T^2_3:  \qquad   \tan \pi \tilde{\varphi}^a_3 =\frac{M^a_3}{n^a_3}r, 
\end{equation}
with $ M^a_3 \equiv m^a_3 +b n^a_3 $, the necessary supersymmetry condition in terms of wrapping numbers is
\begin{eqnarray}
{\bf AAa/b} & \qquad&  Z_a n^a_3 + Y_a M^a_3 r =0, \nonumber \\
{\bf ABa/b} & \qquad&  \left(Z_a - Y_a \right) n^a_3 +\left(Y_a+Z_a \right) M^a_3 r =0 \nonumber,\\
{\bf BBa/b} & \qquad&  -Y_a n^a_3 +Z_a M^a_3 r =0. \label{Eq:SUSYcond_1}
\end{eqnarray}
Since the tangent is defined modulo $\pi$, (\ref{Eq:SUSYcond_1}) does not distinguish between 
D6-branes and anti-D6-branes. The sufficient condition for supersymmetric D$6_a$-branes is the 
correct combination of signs for $(Y_a, Z_a)$ and $(n^a_3,M^a_3)$ such that all contributions to the
RR tadpoles are positive, see eq.~(\ref{Eq:RRtadpoles}) below. 

The O6-planes lie on the fixed planes of ${\cal R}\Theta^k\omega^l$ and are explicitly given in terms of 
wrapping numbers and 3-cycles
in table~3. Since e.g. on an {\bf A} torus, two parallel vertical or horizontal planes passing through the origin 
and displaced by 
$\frac{1}{2}\pi_{2k-1}$ or $\frac{1}{2}\pi_{2k}$  are invariant under ${\cal R}\Theta^{2k}\omega^l$,
in the last column the number of parallel O6-planes is listed. The over-all
cycle $\Pi_{O6}$ is the sum over all possible fixed planes weighted by their multiplicities. 
\begin{table}[h]
\tbl{O6-planes for $T^6/({\mathbb Z}_4 \times {\mathbb Z}_2 \times \Omega
  {\cal R})$}
{\begin{tabular}{@{}cccccc@{}} \toprule 
lattice  & $(n_1,m_1;n_2,m_2;n_3,m_3)$ & $Y$ & $Z$ & cycle 
& $T_1 \times T_2 \times T_3$ \\ \colrule
{\bf AAa/b} & $(1,0;1,0;\frac{1}{1-b},\frac{-b}{1-b})$ & 1 & 0 
& $\frac{1}{1-b}\left[\rho_1 -b\rho_3 \right]$ & $2 \times 2 \times  2(1-b)$\\
 & $(1,1;1,-1;\frac{1}{1-b},\frac{-b}{1-b})$ & 2 & 0 
& $\frac{2}{1-b}\left[\rho_1 -b\rho_3 \right]$  & $1 \times 1  \times 2(1-b)$ \\
 & $(1,0;0,1;0,-1)$ & 0 & 1 & $-\rho_4$ & $2 \times 2 \times 2(1-b)$\\
 & $(1,1;1,1;0,-1)$ & 0 & 2 & $-2\rho_4$ & $1 \times 1  \times 2(1-b)$\\\colrule
{\bf ABa/b} & $(1,0;1,1;\frac{1}{1-b},\frac{-b}{1-b})$ & 1 & 1 
& $\frac{1}{1-b}\left[\rho_1+\rho_2-b(\rho_3+\rho_4) \right]$ & $2 \times 1 \times  2(1-b)$\\
 & $(1,1;1,0;\frac{1}{1-b},\frac{-b}{1-b})$ & 1 & 1 
&  $\frac{1}{1-b}\left[\rho_1+\rho_2-b(\rho_3+\rho_4) \right]$ & $1 \times 2  \times 2(1-b)$ \\
 & $(1,0;-1,1;0,-1)$ & -1 & 1 & $\rho_3 - \rho_4$ & $2 \times 1 \times  2(1-b)$\\
 & $(1,1;0,1;0,-1)$ & -1 & 1 &  $\rho_3 - \rho_4$ & $1 \times 2  \times 2(1-b)$\\\colrule
{\bf BBa/b}  & $(0,1;1,0;\frac{1}{1-b},\frac{-b}{1-b})$ & 0 & 1 & $\frac{1}{1-b}\left[\rho_2
   -b\rho_4 \right]$ & $2 \times 2 \times 2(1-b)$\\
& $(1,1;1,1;\frac{1}{1-b},\frac{-b}{1-b})$ & 0 & 2 
&  $\frac{2}{1-b}\left[\rho_2 -b\rho_4 \right]$& $1 \times 1  \times 2(1-b)$\\
& $(0,1;0,1;0,-1)$ & -1 & 0 & $\rho_3$ & $2 \times 2 \times 2(1-b)$\\
 & $(1,1;-1,1;0,-1)$ & -2 & 0 & $2\rho_3$& $1 \times 1  \times 2(1-b)$\\
\botrule
\end{tabular}}
\end{table}

Taking into account the
sublety that, effectively, the O6-planes wrap the short cycles
$\rho_i$ while $N_a$ D6-branes wrapping the long cycles $\rho_i^{\prime}$ support the gauge group $U(N_a)$, 
the RR tadpole cancellation condition takes the form
\begin{eqnarray}
{\bf AAa/b}: \;\;
&\sum_a& N_a \left[Y_an^a_3\left(\rho_1-b\rho_3\right)+Z_a M^a_3\rho_4 \right] 
=12 \left[\left(\rho_1-b\rho_3\right)-(1-b)\rho_4 \right],  \nonumber\\
{\bf ABa/b}: \;\;
&\sum_a& N_a  \left[\left(Y_a+Z_a \right)n^a_3\left(\rho_1+\rho_2-b(\rho_3+\rho_4)\right)+\left(Y_a-Z_a \right)
M^a_3 \left(\rho_3-\rho_4\right)   \right] \nonumber\\
&=& 16  \left[\left(\rho_1+\rho_2\right)-b\left(\rho_3+\rho_4\right) +(1-b)\left(\rho_3-\rho_4\right)\right], \nonumber\\
{\bf BBa/b}:\;\;
&\sum_a& N_a  \left[Z_an^a_3\left(\rho_2-b\rho_4\right)+Y_a M^a_3 \rho_3 \right] 
= 12\left[\left(\rho_2-b\rho_4\right)+(1-b)\rho_3  \right]. \; \label{Eq:RRtadpoles}
\end{eqnarray}
For the explicit computation of the RR tadpole cancellation conditions via open string 1-loop amplitudes and worldsheet duality see\cite{Forste:2000hx,Honecker:2003vq}.

The resolution of the cycles in terms of ${\cal R}$ even and odd components is given in table~4 with 
$\Pi_i^{+} \circ \Pi_j^{-}=-\frac{\kappa}{1-b}\delta_{ij}$ and $\kappa=1$ for {\bf AAa/b}, {\bf BBa/b} and 
$\kappa=2$ for {\bf ABa/b}.
\begin{table}[h]
\tbl{${\cal R}$ even and odd cycles for $T^6/({\mathbb Z}_4 \times {\mathbb Z}_2 \times \Omega {\cal R})$}
{\begin{tabular}{@{}ccccc@{}} \toprule
lattice & $\Pi_1^{+}$ & $\Pi_2^{+}$ & $\Pi_1^{-}$ & $\Pi_2^{-}$ \\ \colrule
{\bf AAa/b} &  $\frac{1}{1-b} \left[\rho_1-b\rho_3 \right] $ &  $\rho_4 $ &   $\rho_3 $
&  $-\frac{1}{1-b} \left[\rho_2-b\rho_4 \right] $\\
{\bf ABa/b} &  $\frac{1}{1-b} \left[\rho_1+\rho_2-b(\rho_3+\rho_4) \right]$ & $\rho_3-\rho_4$
& $\rho_3+\rho_4$  &  $-\frac{1}{1-b} \left[\rho_1-\rho_2-b(\rho_3-\rho_4) \right]$ \\
{\bf BBa/b} &  $\frac{1}{1-b} \left[\rho_2-b\rho_4 \right]$ & $\rho_3$
 & $\rho_4$ & $-\frac{1}{1-b} \left[\rho_1-b\rho_3 \right]$ \\\botrule
\end{tabular}}
\end{table}
The conclusion that two RR scalars participate in the generalized Green
Schwarz mechanism and thus at most two $U(1)$ factors can acquire a mass
is confirmed by 
the explicit calculation of the closed string spectrum displayed in table~5. The untwisted sector provides
two real RR scalars. Any other RR sector contains either vectors or no massless states. In table~5,
the bosonic degrees of freedom (d.o.f.) in terms of real scalars and vectors are listed. The fermionic d.o.f. follow from
supersymmetry. It can be easily checked that the number of closed string
chiral and vector multiplets is indeed $h_{1,1}+h_{2,1}=62$.
\begin{table}[h]
\tbl{Closed string spectrum for $T^6/({\mathbb Z}_4 \times {\mathbb Z}_2 \times
  \Omega {\cal R})$}
 {\begin{tabular}{@{}cc@{}} \toprule
untwisted & NSNS: Graviton + Dilaton + 7 scalars; RR: Axion + 1 scalar   \\ \colrule
$\theta+\theta^3$ & NSNS: 16 scalars, RR: --- \\\colrule
$\theta^2$  & {\bf AAa/b, BBa/b:} NSNS: 20 scalars, RR: ---
{\bf ABa/b:} NSNS: 18 scalars, RR: 1 vector \\\colrule
$\omega$  & {\bf AAa, ABa, BBa:} NSNS: 24 s  RR: --- {\bf AAb, ABb, BBb:}  NSNS: 18 s RR: 3 v   \\\colrule 
$\theta\omega+\theta^3\omega$ & {\bf AAa, ABa, BBa:} NSNS: 32 s RR: ---  {\bf AAb, ABb, BBb:}  NSNS: 24 s RR: 4 v    \\\colrule
$\theta^2\omega$  &  {\bf AAa, ABa, BBa:} NSNS: 24 s RR: --- {\bf AAb, ABb, BBb:}  NSNS: 18 s RR: 3 v   \\\botrule
\end{tabular}}
\end{table}

From the cycles~(\ref{cycle_YZ}), their ${\cal R}$ images~(\ref{Eq:Rimagecycles}), the supersymmetry condition~(\ref{Eq:SUSYcond_1}), 
RR tadpole cancellation~(\ref{Eq:RRtadpoles})
and coefficients for the Green Schwarz couplings, any massless chiral spectrum and its physical $U(1)$ factors
can be computed.

In the following section, some specific models are discussed.


\subsection{Chiral spectra: A no-go theorem for three generations}\label{Sec:No_3_gen}

The main goal of four dimensional string compactifications is to find stable  vacua with 
the gauge group and chiral matter content of the standard model or, e.g. an
$SU(5)$ 
or Pati-Salam GUT model. 
While engineering the required gauge group is comparatively easy, obtaining three quark generations 
is a complicated, in some scenarios even impossible task.

Three different phenomenologically appealing scenarios with a supersymmetric visible sector are conceivable 
in the framework of D6-branes intersecting at angles:
\begin{romanlist}
\item The gauge group contains $SU(3) \times SU(2) \times U(1)$. All left- and right-handed quarks 
are realized as bifundamental representations. Pati-Salam and left-right symmetric models also belong to this 
category. 
\item  The gauge group contains $SU(3) \times SU(2)$. The left-handed quarks transform as bifundamentals, 
whereas the right-handed quarks are realized as antisymmetric representations of $SU(3)$. 
\item  The gauge group contains $SU(5) \times U(1)$. The left handed quarks sit in three antisymmetric representations
of $SU(5)$, and three bifundamentals provide the remaining right-handed quarks.    
\end{romanlist}
The three scenarios have to fulfill the following minimal requirements in order to 
provide stable\cite{Rabadan:2001mt,Cremades:2002te} genuine three generation models:
\begin{itemlist}
 \item Supersymmetry of the visible sector.
 \item Three intersections of the QCD-stack with a second stack.
 \item No chiral fermion in the symmetric representation of the QCD-stack.
\end{itemlist} 
The intersection numbers of factorizable 3-cycles
and their ${\cal R}$ images can be computed from~(\ref{IntersectionMatrix}), (\ref{cycle_YZ}) and (\ref{Eq:Rimagecycles}),
\begin{eqnarray}
\Pi_a \circ \Pi_b &=& \left(Y_aY_b+Z_aZ_b \right)\left( -n^a_3M^b_3+M^a_3n^b_3 \right),\nonumber\\
\Pi_a \circ \Pi_{b'} &=&  \left\{ \begin{array}{ll} 
\left(Y_aY_b-Z_aZ_b \right)\left( n^a_3M^b_3+M^a_3n^b_3 \right)  & {\bf AAa/b},\nonumber\\
\left(Y_aZ_b+Z_aY_b \right)\left( n^a_3M^b_3+M^a_3n^b_3 \right) & {\bf ABa/b},\nonumber\\
\left(-Y_aY_b+Z_aZ_b \right)\left( n^a_3M^b_3+M^a_3n^b_3 \right) & {\bf BBa/b}.
\end{array}\right.
\end{eqnarray}
The net number of fundamental representations of $U(N_a)$  is therefore given by  
\begin{equation}\label{Eq:Net_No_Generations}
\Pi_a \circ \left( \Pi_b +  \Pi_{b'} \right)= \left\{ \begin{array}{ll}
2\left[Y_aY_bM^a_3n^b_3-Z_aZ_bn^a_3M^b_3\right]  & {\bf AAa/b},\\
(Y_a+Z_a)(Y_b+Z_b)M^a_3n^b_3-(Y_a-Z_a)(Y_b-Z_b)n^a_3M^b_3   & {\bf ABa/b},\\
2\left[-Y_aY_bn^a_3M^b_3+Z_aZ_bM^a_3n^b_3\right] & {\bf BBa/b}.
\end{array}\right.
\end{equation}
For the untilted $T^2_3$, $b=0$ implies that $M^c_3=m^c_3 \in {\mathbb Z}$, and it is evident that 
the {\bf AAa} and {\bf BBa} orientations can only provide even numbers of
fundamental representations of  $U(N_a)$. This observation is independent of supersymmetry.

If stacks of  D$6_b$-branes are realized  on top the O6-planes, the gauge group is $Sp(2N_b)$, 
and the number of bifundamental representations is given by \mbox{$\Pi_a \circ \Pi_b$}. In table~3, the wrapping 
numbers for the different O6-planes are given. If the corresponding cycles are labeled by $\Pi_{b1}, \Pi_{b2}$ 
for  $n_3 \neq 0, M_3 =0$ and $\Pi_{c1}, \Pi_{c2}$ for $n_3 = 0, M_3 \neq 0$, the intersection numbers of an
arbitrary factorizable cycle $\Pi_a$ with the ${\cal R}$ invariant ones are 
\begin{equation}\label{Eq:Net_Sp_Generations}
\Pi_a \circ \Pi_{bl} = \frac{1}{1-b}M^a_3 \left\{ \begin{array}{l}
l Y_a, \\ (Y_a+Z_a), \\ l Z_a, \end{array}\right.
\qquad 
\Pi_a \circ \Pi_{ck} = n^a_3 \left\{ \begin{array}{ll}
k Z_a  & {\bf AAa/b},\\ (Z_a - Y_a)  & {\bf ABa/b},\\ - k Y_a  & {\bf BBa/b}.
 \end{array}\right.
\end{equation}
For $k,l=2$ and the {\bf AAa/b}, {\bf BBa/b} lattices, odd numbers of generations are not possible. 
As in the discussion of~(\ref{Eq:Net_No_Generations}), the statement is independent of supersymmetry.

The number of (anti)symmetric chiral representations is computed from
\begin{eqnarray}
\Pi_a \circ \Pi_{a'} &=&  2n^a_3M^a_3\left\{ \begin{array}{ll}
\left(Y_a^2-Z_a^2 \right) & {\bf AAa/b}, \\  2Y_aZ_a& {\bf ABa/b},  \\ \left(Z_a^2-Y_a^2\right)& {\bf BBa/b},
 \end{array}\right. \nonumber\\
\Pi_a \circ  \Pi_{O6} &=& 2 \left\{ \begin{array}{ll}
3 \left[Y_a M^a_3+(1-b)Z_a n^a_3\right] & {\bf AAa/b},\\  
2\left[\left(Y_a+Z_a\right)M^a_3-(1-b)\left(Y_a-Z_a\right)n^a_3 \right] & {\bf ABa/b},\\
3\left[-(1-b)Y_an^a_3+Z_a M^a_3 \right] & {\bf BBa/b},
\end{array}\right.\label{Eq:AntiSym_Generations}
\end{eqnarray}
and only for $\Pi_a \circ \Pi_{a'}=\Pi_a \circ  \Pi_{O6}$, the spectrum is free of chiral symmetric states.
By~(\ref{Eq:SUSYcond_1}), at the special radii $r=\frac{1}{1-b}$ all supersymmetric 3-cycles fulfill
$\Pi_a \circ  \Pi_{O6}=0$. This means that the number of antisymmetric and symmetric representations are
zero simultaneously and only scenario (i) is potentially accessible. 

Formally, $\Pi_a \circ \Pi_{a'}=\Pi_a \circ  \Pi_{O6}$, $\Pi_a \circ \left( \Pi_b + \Pi_{b'} \right)= \pm 3$ 
(or $\Pi_a \circ \Pi_{bl/ck}=\pm 3$ due to $SU(2) \equiv Sp(2)$) and the supersymmetry condition~(\ref{Eq:SUSYcond_1})
can be solved for  $(n^a_3,m^a_3)$, $(Y_a,Z_a)$, $b$  and $r$. 
It turns out that the solutions require a square $T^2_3$, i.e. $r=\frac{1}{1-b}$, and e.g. on {\bf BBA/B} 
$(n^a_3,m^a_3)=(1,-1)$, $(Y_a,Z_a)=(-3,3)$  up to exchanging $\Pi_a$ with its ${\cal R}$ image $\Pi_{a'}$.
However, the expansion~(\ref{Eq:Def_YZ}) in terms of wrapping numbers $(n^a_k,m^a_k)$ which have to be coprime 
per $T^2_k$ ($k=1,2$) does not have any solution to $|Y_a|=|Z_a|=3$. The computation for the {\bf AAa/b} and {\bf ABa/b} orientations 
is analogous. In particular, all results for {\bf AAa/b} are obtained from those of {\bf BBa/b} by substituting 
$(Y_a, Z_a) \rightarrow (-Z_a, Y_a)$.

In summary, on the $T^6/({\mathbb Z}_4 \times {\mathbb Z}_2 \times \Omega{\cal R})$ orientifold no genuine three generation models 
on factorizable 3-cycles fulfilling the requirements stated at the beginning
of this section exist. In the next section, we present instead
a 2+1 generation model.

\subsection{A 2+1 generation model with brane recombination}

In~\cite{Honecker:2003vq,Honecker:2003vw} a 2+1 generation model on the {\bf ABB} lattice was presented which 
contains a breaking $SU(3)^2 \rightarrow SU(3)_{diag}$ and thereby converts the cycle which the QCD-stack wraps into 
a non-factorizable one. The $SU(2)_L \times SU(2)_R$ stacks remain factorizable, allowing for the 
computation of all possible Higgs-multiplet candidates from the non-chiral sector.

In this section, we present a similar construction on the {\bf BBB} lattice. As in~\cite{Honecker:2003vq,Honecker:2003vw},
two non-trivial kinds of D$6_{aj}$-branes with angles 
$(\tilde{\varphi}_1,\tilde{\varphi}_2,\tilde{\varphi}_3)\pi=(1/4,0,-1/4)\pi$ and $(0,1/4,-1/4)\pi$ w.r.t. the ${\cal R}$ 
invariant axis together with the D$6_{bl}$- and D$6_{ck}$-branes on top of some O6-planes are needed. 
The set-up is given in table~6.
\begin{table}[h]
\tbl{D6-brane configuration for a 2+1 generation model with $b=\frac{1}{2}$ and $r=2$}
{\begin{tabular}{@{}cccccc@{}} \toprule
D6-brane & $N$ & $(n_1,m_1;n_2,m_2;n_3,m_3)$ & $Y$ & $Z$ & cycle \\\colrule
$a_1$ & 3 & (0,1;1,1;1, -1) & -1 & 1 &  $\left[-\rho_1+\rho_2-(-\rho_3+\rho_4) \right]$\\
$a_2^{(i)}$, $i=1,2,3$ &  1 & (1,1;0,1;1, -1) & -1 & 1 &  $\left[-\rho_1+\rho_2-(-\rho_3+\rho_4) \right]$\\
$b_1$ & 1 & (0,1;1,0;2,-1) & 0 & 1 & $2\left[\rho_2  -b\rho_4 \right]$ \\
$b_2$ & 1 & (1,1;1,1;2,-1) & 0 & 2 & $4\left[\rho_2 -b\rho_4 \right]$\\
$c_1$ & 1 & (0,1;0,1;0,-1) & -1 & 0 & $\rho_3$ \\
$c_2$ & 1 & (1,1;-1,1;0,-1) & -2 & 0 & $2\rho_3$\\\botrule
\end{tabular}}
\end{table}
The resulting gauge group is 
\begin{equation}
U(3)_{a1} \times \left( U(1)_{a2} \right)^3 \times SU(2)_{b1}  \times SU(2)_{b2} \times SU(2)_{c1}  \times SU(2)_{c2}
\end{equation}
with the identification $Sp(2) \equiv SU(2)$. The three stacks $a_2^{(i)}$ wrap the same 
cycles and are parallely displaced on some 2-torus, e.g. on $T^2_3$. 

The massless $U(1)$ factors can be computed using~(\ref{Eq:masslessAbelian}) and table~4. It turns out, that 
apart from $B-L$, which occurs in all known left-right symmetric models with D6-branes at angles, 
two additional $U(1)$ factors remain massless but without
obvious physical interpretation,
\begin{eqnarray}
Q_{B-L} &=& \frac{1}{3} Q_{a1} - Q_{a2}^{(1)}, \nonumber\\
Q' &=& Q_{a2}^{(2)} - Q_{a2}^{(3)},\nonumber\\
Q'' &=& \frac{3}{5}Q_{a1} +\frac{1}{5}Q_{a2}^{(1)}-Q_{a2}^{(2)} - Q_{a2}^{(3)}.\label{Eq:Model_massless_U1}
\end{eqnarray}
The fourth $U(1)$ factor acquires a mass through its non-vanishing coupling
to the $B^i_2$.
The gauge group of the set-up with factorizable 3-cycles therefore is 
\begin{equation}
SU(3)_{a1} \times \underbrace{SU(2)_{b1}  \times SU(2)_{b2}} \times \underbrace{SU(2)_{c1}  \times SU(2)_{c2}} \times U(1)_{B-L}
\times U(1)' \times U(1)'' \times U(1)_{m}. \nonumber
\end{equation}
The chiral spectrum is given in table~7.
\begin{table}[h]
\tbl{Chiral spectrum of the 2+1 generation model} 
{\begin{tabular}{@{}cccc|ccc@{}} \toprule
Sector & rep & $B-L$ & particle $\;\;$& $\;\;$ Sector & rep & $B-L$ \\ \colrule
$a_1 b_1$ & $(\overline{{\bf 3}}_{a1},{\bf 2}_{b1})$ & -1/3 & $Q_R$  & $a_2^{(2)} b_1$ & $(\overline{{\bf 1}}_{a2^{(2)}},{\bf 2}_{b1})$  & 0\\
$a_1 b_2$ & $2(\overline{{\bf 3}}_{a1},{\bf 2}_{b2})$ & -1/3 & $Q_R$ & $a_2^{(2)} b_2$ & $2(\overline{{\bf 1}}_{a2^{(2)}},{\bf 2}_{b2})$ & 0 \\ 
$a_1 c_1$ & $({\bf 3}_{a1},{\bf 2}_{c1})$ & 1/3 & $Q_L$ & $a_2^{(2)} c_1$ &  $({\bf 1}_{a2^{(2)}},{\bf 2}_{c1})$ & 0\\
$a_1 c_2$ & $2({\bf 3}_{a1},{\bf 2}_{c2})$ & 1/3 & $Q_L$ & $a_2^{(2)} c_2$ & $2({\bf 1}_{a2^{(2)}},{\bf 2}_{c2})$  & 0\\\hline
$a_2^{(1)} b_1$ & $(\overline{{\bf 1}}_{a2^{(1)}},{\bf 2}_{b1})$ & 1 & $E_R, N_R$ & $a_2^{(3)} b_1$ 
& $(\overline{{\bf 1}}_{a2^{(3)}},{\bf 2}_{b1})$  & 0\\
$a_2^{(1)} b_2$ & $2(\overline{{\bf 1}}_{a2^{(1)}},{\bf 2}_{b2})$ & 1 & $E_R, N_R$ & a$_2^{(3)} b_2$ 
& $2(\overline{{\bf 1}}_{a2^{(3)}},{\bf 2}_{b2})$  & 0\\
$a_2^{(1)} c_1$ &  $({\bf 1}_{a2^{(1)}},{\bf 2}_{c1})$ & -1 & $L$ & $a_2^{(3)} c_1$ &  $({\bf 1}_{a2^{(3)}},{\bf 2}_{c1})$  & 0\\
$a_2^{(1)} c_2$ & $2({\bf 1}_{a2^{(1)}},{\bf 2}_{c2})$ & -1 & $L$ & $a_2^{(3)} c_2$ & $2({\bf 1}_{a2^{(3)}},{\bf 2}_{c2})$ & 0 \\\botrule
\end{tabular}}
\end{table}
In order to shorten the notation, a ${\bf 3}_{-1}$ of $SU(3)_{a1} \times U(1)_{a1}$ is denoted as $\overline{\bf 3}_{a1}$ and 
$\overline{{\bf 1}}_{a2^{(1)}}$ denotes charge $-1$ under $U(1)_{a2}^{(1)}$. 
The particles on the left side of the table have the quantum numbers of the left-right symmetric standard model provided that two 
brane recombinations occur, namely $SU(2)_{b1}  \times SU(2)_{b2} \rightarrow SU(2)_R$ and $SU(2)_{c1}  \times SU(2)_{c2} \rightarrow SU(2)_L$.
In both cases, $b_1$ and $(\Theta^k b_2)$ as well as $c_1$ and $(\Theta^k c_2)$ are parallel on $T^2_3$ and intersect once 
on $T^2_1 \times T^2_2$ for $k=0,1$ each. The intersections provide two hypermultiplets in $({\bf 2}_{b1},{\bf 2}_{b2})$ and 
$({\bf 2}_{c1},{\bf 2}_{c2})$ which are required for D- and F-flatness of the brane recombination process.
Since the recombined branes still factorize into 2-cycles times 1-cycles on $T^4 \times T^2_3$, the recombined stack $B=b_1+b_2$ can be displaced from 
the ${\cal R}$ invariant position on $T^2_3$. In this way, the electro weak breaking $SU(2)_R \times U(1)_{B-L} \rightarrow U(1)_Y \times U(1)'''$ can be
realized.

The states listed on the right side of the table constitute additional vector-like matter.

A similar construction is possible on the {\bf AAB} lattice as is immediately evident from the symmetry of the 
formulae under the exchange $(Y_a, Z_a) \rightarrow (-Z_a, Y_a)$.

In constrast to the previously presented model on the {\bf ABB} lattice, the construction is not based on
a Pati-Salam group $SU(4)$, and the non-factorizable branes arise from the
`leptonic' and `right' stacks. The lengths of the cycles $\Pi_{a1}$ and
$\Pi_{a2}$, however, are identical as in any Pati-Salam construction and thus fulfill a minimal requirement 
on a possible gauge unification.\cite{Blumenhagen:2003jy}

\section{Genuine three generation models on isotropic tori}

In section~\ref{Sec:No_3_gen}, it has been  shown that on the $T^6/({\mathbb
  Z}_4 \times {\mathbb Z}_2 \times \Omega {\cal R})$ orientifold,
phenomenologically interesting supersymmetric genuine 3-generation models are not accessible. The analysis
includes only the conditions for supersymmetry, no symmetric representations on the QCD-stack and three quark generations.
If these conditions were fulfilled, the next requirement would be that the QCD-stack did not intersect with 
any of the remaining stacks of the model besides from those providing the left- and right-handed quarks.
In the following, we briefly discuss how phenomenological model building with D6-branes at angles in other
  orbifold backgrounds is affected by these restrictions.

The remaining symmetric orbifolds which preserve ${\cal N}=1$ supersymmetry and have factorizable tori and 3-cycles 
fall into two classes. The $T^6/({\mathbb Z}_2 \times {\mathbb Z}_2)$\cite{Berkooz:1996dw,Forste:2000hx,Cvetic:2001tj,Cvetic:2001nr}, $T^6/{\mathbb Z}_3$\cite{Blumenhagen:1999ev,Pradisi:1999ii,Blumenhagen:2001te} and 
$T^6/({\mathbb Z}_3 \times {\mathbb Z}_3)$\cite{Forste:2000hx} orbifolds have only bulk cycles and ${\cal N}=4$ gauge multiplets, 
whereas fractional 3-cycles and ${\cal N}=2$ gauge multiplets occur for $T^6/{\mathbb Z}_N$ ($N=4,6,6'$)\cite{Blumenhagen:1999ev,Blumenhagen:2002gw,Honecker:2004kb} and 
$T^6/({\mathbb Z}_6 \times {\mathbb Z}_3)$\cite{Forste:2000hx,Honecker:2004kb}. 

\subsection{Orbifolds with factorizable O6-planes and bulk cycles only}

The $T^6/({\mathbb Z}_2 \times {\mathbb Z}_2)$ case has been studied in great detail.
The ${\mathbb Z}_2$ actions do not constrain the ratio of radii on any 2-torus, and there is a rich variety of models
which in general require anisotropic compact dimensions.
However, also on this orbifold, there is not any known model fulfilling all
requirements mentioned in section~\ref{Sec:No_3_gen}
plus vanishing intersections of the QCD-stack with any further stack, 
see e.g.~\cite{Cvetic:2002pj,Cvetic:2003xs,Cvetic:2004ui}, and the exotic chirals have to confine into composite matter 
fields due to a strongly coupled hidden sector\cite{Cvetic:2002qa} which
can also lead to a dynamical supersymmetry breaking.\cite{Cvetic:2003yd}

The $T^6/{\mathbb Z}_3$ and $T^6/({\mathbb Z}_3 \times {\mathbb Z}_3)$
orbifold are not suitable for 
supersymmetric model building since only two linearly independent bulk cycles exist. Supersymmetry projects onto 
one specific bulk cycle, and all intersection numbers among D6-branes and with the O6-planes vanish automatically.
Therefore, no chiral fermions arise in supersymmetric set-ups.

\subsection{Orbifolds with factorizable O6-planes and fractional cycles}

For the $T^6/{\mathbb Z}_4$ case, the RR tadpole cancellation conditions and supersymmetric fractional cycles were 
computed and a 1+1+1 generation model has been presented~\cite{Blumenhagen:2002gw}.
The shape of the tori is identical to the 
$T^6/({\mathbb Z}_4 \times {\mathbb Z}_2)$ case, but the exceptional cycles change the
pattern of possible intersection numbers. The examination of this pattern shows that 
supersymmetric genuine three generation models are not accessible. 

The $T^6/({\mathbb Z}_6 \times {\mathbb Z}_3)$ orbifold has,\cite{Honecker:2004kb} up to normalization, 
the same bulk 3-cycles as $T^6/{\mathbb Z}_3$
plus two exceptional 3-cycles stuck at the ${\mathbb Z}_2$ fixed points. 
The geometric interpretation of the loop channel amplitudes enforces 
a specific combination of the supersymmetric bulk 3-cycle with exceptional
ones, not leaving enough freedom for phenomenological model building.
The $T^6/{\mathbb Z}_6$ and $T^6/{\mathbb Z}_6^{\prime}$ orbifolds have enough bulk and exceptional 3-cycles
to be of phenomenological interest. While the latter case is currently under investigation\cite{HoneckerOtt:work}, 
the former has already been examined systematically\cite{Honecker:2004kb}. 
It turns out that in this background
genuine  three generation models exist which fulfill all the requirements mentioned above. 
As observed first for toroidal compactifications\cite{Ibanez:2001nd}, the RR tadpole
cancellation conditions are stronger than the conditions on vanishing gauge
anomalies. Therefore, all models constructed in\cite{Honecker:2004kb} have
a non-minimal Higgs sector. In particular, one obtains
exactly the chiral spectrum of the supersymmetric
standard model with three Higgs  generations and isotropic compact dimensions.
In the remainder of this section, we briefly comment on the construction of
this supersymmetric standard model on fractional branes in the $T^6/{\mathbb Z}_6$ case.

\subsubsection{Generalities on fractional branes}

In order to clarify the role of the exceptional cycles, let us first look at 
six dimensional IIB orientifolds on ${\mathbb R}^{1,5} \times T^4/({\mathbb
  Z}_N \times \Omega {\cal R})$. The closed string spectrum of the IIB theory 
contains\cite{Douglas:1996sw} in the $k^{th}$ twisted sector a NSNS triplet of scalars $\vec{\chi}_k$  
associated to the metric moduli (complex and K\"ahler deformations) and a NSNS scalar
$b^{(0)}_k = \int_{e^{(k)}}{}^{(10)}B_2$ per fixed point. The RR sector
contributes a RR scalar $\phi_k  = \int_{e^{(k)}}{}^{(10)}C_2$ and a 2-form
$C^k_2= \int_{e^{(k)}}{}^{(10)}C_4$ per fixed point. The fermionic
superpartners arise from the R-NS and NS-R sectors.
 
The orientifold projection $\Omega$ identifies the $k^{th}$ twisted sector
with its inverse \mbox{$(N-k)^{th}$} sector.
For $N$ odd, the counting of the twisted
closed string states comprises the sectors $k=1,\ldots,[N/2]$ and $k=[N/2]+1,
\ldots, N-1$ contains no further physical d.o.f.. In terms of the Fourier
transformed formulation, this means that exceptional cycles from inverse
sectors are identified, $e^{(k)} \stackrel{\Omega}{\rightarrow} \pm e^{(N-k)}$.
For $N$ odd, an additional sign in the projection corresponds to a redefinition of 
the inverse cycle and does not affect the closed string spectrum.
For $N=2M$, the same reasoning applies to all sectors with $k \neq M$, whereas 
the contribution to the closed string spectrum from  ${\mathbb Z}_2$ twisted sectors
depends on the sign which the exceptional cycle picks up.

The general situation is different for the projection $\Omega{\cal R}$.
In this case, the orientifold projection preserves each twist sector
separately, $e^{(k)} \stackrel{\Omega{\cal R}}{\rightarrow} e^{(k)}$, and the
explicit computation of the closed string spectra\cite{Blumenhagen:1999md,Pradisi:1999ii} 
shows that if the orbifold fixed point at which the cycle $e^{(k)}$ is located is also 
invariant under ${\cal R}$, one hypermultiplet per twist sector emerges, while 
orbifold fixed points which form pairs under ${\cal R}$ support a hyper and a tensor
multiplet.

The O7-planes wrap the factorizable bulk cycles which are invariant under some 
element $\Omega{\cal R} {\mathbb Z}_k$ of the orientifold group. However, in general 
not the bulk cycles but rather linear
combinations of bulk and exceptional 2-cycles of the form
\begin{equation}\label{Eq:FracCycles}
\Pi_a^{frac}=\frac{1}{2}\Pi_a^{bulk} +\frac{(-1)^{\tau_0}}{2}
\left(e_{ik}+(-1)^{\tau_1}e_{jk}+(-1)^{\tau_2}e_{il}+(-1)^{\tau_1+\tau_2}e_{jl}  \right)
\end{equation}
provide an unimodular basis for the lattice on $T^4/{\mathbb Z}_2$
where $i,j \in T^2_1$ and $k,l \in T^2_2$ label the fixed points the bulk cycle passes through. 
$\tau_0=0,1$  corresponds to the eigenvalues $\pm 1$ of ${\mathbb Z}_2$, and $\tau_k =0,1$
denotes a  discrete Wilson line on $T^2_k$ ($k=1,2$).
Any bulk cycle which passes through the orbifold fixed points can be
  decomposed into its fractional components, 
$\Pi_a^{bulk}=\frac{1}{2}\left(\Pi_a^{bulk}+\Pi_a^{ex}
  \right)+\frac{1}{2}\left(\Pi_a^{bulk}-\Pi_a^{ex} \right)$. While $N_a$
  D7-branes on a bulk cycle away from the orbifold point provide the gauge
  group $U(N_a)_{diag}$, on the fixed points (at the `enhancon') each kind of $N_a$
  opposite fractional D7-branes provides the gauge group $U(N_a)$, i.e. 
$U(N_a)_{diag} \rightarrow U(N_a)^2$.\cite{Bertolini:2003iv} If furthermore the cycles are
their own $\Omega {\cal R}$ images, each $U(N_a)$ is enhanced to $SO(2N_a)$ or $Sp(2N_a)$.

The general line of reasoning carries over to IIA models on $T^6/{\mathbb
  Z}_{2N}$ with D6-branes. In the ${\mathbb Z}_2$ subsector, the theory takes the local form
  $T^4/{\mathbb Z}_2 \times T^2$, and the IIA orientifold is obtained from the
  IIB theory in this section by applying one T-duality
  along the additional 2-torus. The exceptional 3-cycles decompose into
  exceptional 2-cycles times bulk 1-cycles, and the RR fields of the
  effective four dimensional theory are given in~(\ref{Eq:4dRRfields}).
The orbifold invariant bulk and exceptional cycles are computed along the same
  lines as for the $T^6/({\mathbb Z}_4 \times {\mathbb Z}_2)$ case described
  in section~\ref{Subsec:Geometry}, and the fractional cycles are computed in the
  spirit of~(\ref{Eq:FracCycles}).

\subsubsection{Three generations on the $T^6/{\mathbb Z}_6$ orbifold}\label{Z6_model}

The $T^6/({\mathbb Z}_6 \times \Omega {\cal R})$ model with fractional
D6-branes at angles is the first known case with three quark generations and
no additional chiral matter charged under the QCD-stack. 
In this construction, the replication of generations occurs via an intriguing
interplay between discrete Wilson lines and the ${\mathbb Z}_3$ symmetry on
the additional 2-torus, see figure~2. The observation that three families arise naturally 
from ${\mathbb Z}_3$ singularities is known from heterotic compactifications for nearly 
\mbox{20 years}\cite{Ibanez:1987sn}. However, the discrete Wilson lines 
in the model with D6-branes at angles  belong to the ${\mathbb Z}_2$ singularities of the 
orbifold symmetry ${\mathbb Z}_6$.
\begin{figure}[th]
\centerline{\psfig{file=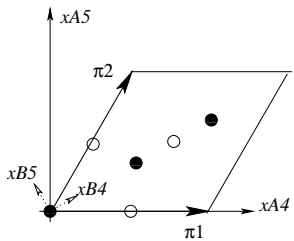,width=1.5in}}
\vspace*{8pt}
\caption{Torus with ${\mathbb Z}_3$ invariance. Filled circles denote points fixed under ${\mathbb Z}_3$. In case of a ${\mathbb Z}_6$ 
symmetry, the 
empty circles are invariant under the ${\mathbb Z}_2$ subgroup. The origin is fixed under the complete orbifold group, 
whereas the fixed points of ${\mathbb Z}_{2},{\mathbb Z}_{3}$ elements are
permuted by the generator of ${\mathbb Z}_6$. As for the square tori of
section~\ref{Subsec:Geometry}, two orientations {\bf A} and {\bf B} are
consistent with the involution ${\cal R}$.}
\end{figure}

The details of the construction can be found in\cite{Honecker:2004kb}. The chiral content
of the supersymmetric standard model can be obtained from five stacks of D6-branes in a great 
variety of geometric settings. The chiral spectrum is insensitive to the specific realization
and listed in table~8. The initial gauge group of the construction 
is $U(3)_a \times U(2)_b \times U(1)_c \times U(1)_d \times U(1)_e$, and after the generalized
Green Schwarz mechanism, the anomaly-free group is 
\begin{equation}
SU(3)_a \times SU(2)_b \times U(1)_{B-L} \times U(1)_c \times U(1)_e \times U(1)^2_{massive}.
\end{equation}
The hypercharge is a linear combination of the massless Abelian groups, \mbox{$Q_Y=\frac{1}{2}(Q_{B-L}+Q_c)$.}
\begin{table}[ht]
\tbl{Chiral spectrum of 5 stack models on $T^6/({\mathbb Z}_6 \times \Omega{\cal R})$}
{\begin{tabular}{@{}c|cc|ccccccc@{}} \toprule
& sector & $SU(3)_a \times SU(2)_b$ & $Q_a$ &
$Q_b$ & $Q_c$ & $Q_d$ & $Q_e$ & $Q_{B-L}$ & $Q_Y$ \\\colrule
$Q_L$ & ab' & $3 \times (\overline{\bf 3},{\bf 2})$ & -1 & -1 & 0 & 0 & 0 & $\frac{1}{3}$ & $\frac{1}{6}$ \\
$U_R$ & ac & $3\times ({\bf 3},1)$ & 1 & 0 & -1 & 0 & 0 & $-\frac{1}{3}$ & $-\frac{2}{3}$ \\
$D_R$  & ac' & $3\times ({\bf 3},1)$ & 1 & 0 & 1 & 0 & 0 & $-\frac{1}{3}$ & $\frac{1}{3}$  \\
$L$ & bd' & $3 \times (1,{\bf 2})$ & 0 & 1 & 0 & 1 & 0 & -1& $-\frac{1}{2}$ \\
$E_R$ & cd & $3 \times (1,1$) & 0 & 0 & 1 & -1 & 0 &1 & 1\\
$N_R$ & cd' & $3 \times (1,1)$ & 0 & 0 & -1 & -1 & 0 & 1&0 \\
& be &  $3 \times (1,{\bf 2})$ & 0 & 1 & 0 & 0 & -1 & 0& 0\\
& be' &  $3 \times (1,{\bf 2})$ & 0 & 1 & 0 & 0 & 1 & 0 & 0\\\hline
\end{tabular}}
\end{table}

The masses of the anomalous $U(1)$ factors and the non-chiral spectrum depend on the
geometric details of the construction. In some cases, it is possible to recombine
the stacks $c$ and $e$ such that the remaining two kinds of chiral multiplets obtain the 
quantum numbers of the Higgs particles. In this case, the Higgs sector is three times the 
one of the minimal supersymmetric standard model as expected from the `anomaly' constraint
on the $U(2)_b$ gauge factor.

The construction of left-right symmetric extensions of the standard model on 
$T^6/({\mathbb Z}_6 \times \Omega{\cal R})$
follows the same lines, and again the Higgs sector is three times the minimal one.

The phenomenology of this class of models is currently under investigation.\cite{HoneckerOtt:work}

\section*{Acknowledgments}

It is a pleasure to thank T.~Ott for the collaboration on the $T^6/({\mathbb Z}_6 \times \Omega{\cal R})$ 
models.\cite{Honecker:2004kb}\\
This work is supported by the RTN European program under contract number HPRN-CT-2000-00148.

\end{document}